# MTS-DVGAN: Anomaly Detection in Cyber-Physical Systems using a Dual Variational Generative Adversarial Network

Haili Sun[a], Yan Huang[b], Lansheng Han[a,*], Cai Fu[a], Hongle Liu[a], Xiang Long[a]

[a] School of Cyber Science and Engineering, Huazhong University of Science and Technology, Wuhan, China

[b] National Key Laboratory of Science and Technology on Multispectral Information Processing, School of Artificial Intelligence and Automation, Huazhong University of Science and Technology, Wuhan, China

## Abstract

Deep generative models are promising in detecting novel cyber-physical attacks, mitigating the vulnerability of Cyber-physical systems (CPSs) without relying on labeled information. Nonetheless, these generative models face challenges in identifying attack behaviors that closely resemble normal data, or deviate from the normal data distribution but are in close proximity to the manifold of the normal cluster in latent space. To tackle this problem, this article proposes a novel unsupervised dual variational generative adversarial model named MST-DVGAN, to perform anomaly detection in multivariate time series data for CPS security. The central concept is to enhance the model's discriminative capability by widening the distinction between reconstructed abnormal samples and their normal counterparts. Specifically, we propose an augmented module by imposing contrastive constraints on the reconstruction process to obtain a more compact embedding. Then, by exploiting the distribution property and modeling the normal patterns of multivariate time series, a variational autoencoder is introduced to force the generative adversarial network (GAN) to generate diverse samples. Furthermore, two augmented loss functions are designed to extract essential characteristics in a self-supervised manner through mutual guidance between the augmented samples and original samples. Finally, a specific feature center loss is introduced for the generator network to enhance its stability. Empirical experiments are conducted on three public datasets, namely SWAT, WADI and NSL_KDD. Comparing with the state-of-the-art methods, the evaluation results show that the proposed MTS-DVGAN is more stable and can achieve consistent performance improvement.

**Keywords:** Cyber-physical systems (CPS), Anomaly detection, Contrastive constraint, Variational generative adversarial network, Industrial security

## 1 INTRODUCTION

With the advent of the Industry 4.0 revolution [1], people are paying more attention to the application of cyber-physical systems (CPSs) in industrial fields, such as intelligent manufacturing, electronic power grids and smart transportation. CPSs are multidimensional complex systems that integrate computing, networking, and physical environments that contain rich networked actuators and sensors [2], generating large volumes of process data in multivariate time series (MTS) [3]. A typical architecture of CPSs is shown in Figure 1, which consists of three layers: the physical layer, the transport layer and the control layer. Since CPSs play a critical role in industrial system automation, any malicious cyber-physical attacks or operation mistakes may cause enormous damage and lead to serious financial loss, even endangering human life [4].

---

[*]Corresponding Author: Lansheng Han; E-mail address: hanlansheng@hust.edu.cn;

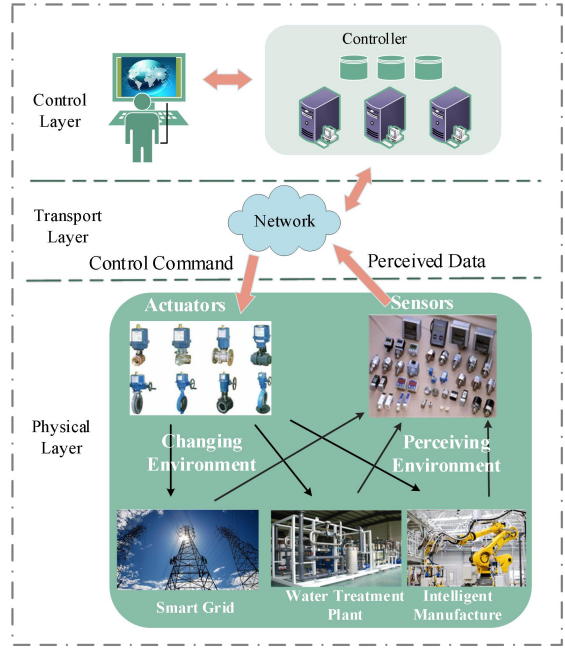

Figure 1: A sketch of the CPS architecture. It consists of three layers: a physical layer, a transport layer and a control layer. The physical layer perceives the environmental (such as smart grid, water treatment and intelligent manufacture) conditions by specific sensors and sends them to the control layer through the transport layer, while the control layer makes decisions based on received environmental data and sends corresponding control commands to the actuators of the physical layer to change the environment.

Anomaly detection [5] is proposed as a promising solution to monitor the working condition of CPSs and alarm immediately once out-of-ordinary behaviors are detected. Those behaviors are described as abnormal activities that deviate from normal patterns of CPSs. These out-of-ordinary behaviors usually deviate from the normal data distribution and are called anomalies, contaminants, intrusions, outliers or failures in various application domains [5]. Figure 2 illustrates two anomalies marked by red solid line rectangles in the multivariate time series of industrial CPSs [6]. Traditionally, to identify anomalies, people manually established a fixed threshold based on expert knowledge for each monitored metric [7] (e.g., CPU utilization, memory usage). However, this work is time and labor consuming due to the exponential growth in the scale and complexity of data in recent years. To address this issue, some anomaly detection approaches have been developed for univariate time series [8], [9], where anomalies are identified based on a single specific metric. However, each metric interacts with others due to the inherent characteristics of CPSs. Therefore, one metric cannot well represent the overall status of the CPSs. Furthermore, it is inefficient to simply combine the detection results of a few univariate time series to form the multivariate time series [10].

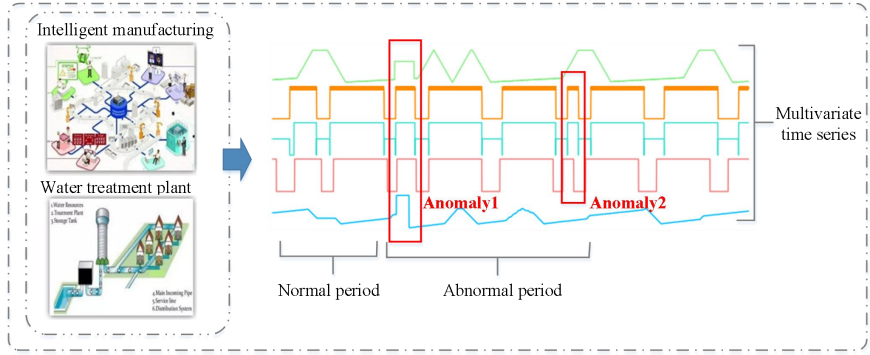

Figure 2: An example of two anomalies (highlighted by a red rectangle) in multivariate time series for industrial CPSs: values of five different sensors are plotted. During the normal period, these values follow a certain pattern, while during the abnormal period, the pattern may break, which indicates that an anomaly has occurred.

In this paper, we focus on identifying anomalies by modeling the multivariate time series of CPSs. Formally, multivariate time series (MTS) are composed of a group of univariate time series (e.g., sensor value), each of which monitors the different parts' status of a complex CPS. Thus, there are rich spatial-temporal correlation dependencies between those MTS, and how to model such dependencies is the key point for detecting anomalies. In general, traditional statistical process control techniques such as Shewhart, EWMA and CUSUM charts were used for quality control [11]. They could discover simple abnormal working states or exception patterns from univariate data of industrial processes. Nevertheless, due to anomaly diversities and the lack of sufficient labeled data, these conventional anomaly detection methods fail to cope with large-scale data produced by modern CPSs, which become increasingly complex and dynamic. To exploit the large amount of multivariate data, many researchers adopt machine learning methods instead of signature-based or specification techniques to detect anomalies [5].

As obtaining inherent labels for anomalies is very challenging, time-consuming and sometimes even impossible, unsupervised learning is regarded as better for anomaly-detection tasks [12]–[14]. However, most existing unsupervised approaches fail to handle multivariate time series with inherent correlation and nonlinearity [15], [16] since they are generally constructed based on transformation and linear projection [17].

To solve the above issue, deep generative models (which identify anomalies based on reconstruction loss) are promising methods for anomaly detection using large-scale multivariate time series; examples include VAE-based [7], [10] and GAN-based methods [16], [18], [19], most of which assume that anomalies deviating from the normal distribution cannot be well reconstructed. However, in the real world, there may exist anomalies that lie near the normal data in distribution whose reconstruction losses are small. Moreover, reconstruction-based approaches may fail to capture particular anomalies that lie far away from the normal data in distribution but near the manifold of the normal class in latent embedding space [20], which makes detecting them quite challenging. Additionally, GAN-based methods have to handle the stabilization and mode collapse issues [21], [22] during the training process.

Therefore, in this work, we propose a novel unsupervised dual variational generative adversarial anomaly detection architecture named MTS-DVGAN, to jointly tackle the aforementioned issues. Specifically, to improve the stability, we first construct a LSTM-based encoder to characterize the spatial and temporal dependencies among multivariate time series and map them into an embedding space. Subsequently, given the representations in the embedding space, a generator with a feature center loss is constructed to well project the representations to reconstructed samples. To enlarge the gap between the reconstruction losses of normal and abnormal data, a augmented module and two augmented losses are proposed to encode essential features of input samples by imposing contrastive constraints on the embedding space in a self-supervised manner through mutual guidance between the augmented samples and input samples. Then, a discriminator is introduced to adversarially boost the generative and discriminative power by predicting whether a sample is generated or not. Finally, the reconstructed loss and discriminative loss are minimized through training and are further utilized to improve the anomaly detection performance.

The advantages of MTS-DVGAN are as follows. First, we map the samples from the real data distribution to latent space via an encoder network to capture the patterns of normal observations. Then, a latent random vector in the latent space is passed to the generator network to reconstruct the original samples and match the feature of the given latent vector with the feature of the original data. With the existence of these anchor spots in latent space, the generator is forced to generate diverse samples. Second, we construct a new optimization objective for the generator to minimize the distance between the feature center of the original data and the generated data instead of employing the traditional binary cross entropy loss. To a certain extent, this loss strategy can alleviate the mode collapse issue in which all generated data move toward a single direction (See Section 4.3 for details). Third, we propose an augmented module to enhance the quality of the latent embedding space features by pulling in the normal samples and pushing away the abnormal samples, which results in a self-supervised mode by guiding the mutual information between the original data and the augmented data. Similarly, another augmented loss is constructed to enhance the quality of the latent embedding space features through the mutual guidance between the latent vector of the augmented data and the original data, which is expected to enhance the discrimination ability of our model.

To verify the performance of our model, we conduct extensive experiments with MTS-DVGAN and state-of-the-art competitors on three public datasets. The results demonstrate that the MTS-DVGAN is more stable and achieves a performance improvement of more than 2.2% in average. Furthermore, the ablation analysis shows the effectiveness of the proposed model. In summary, the main contributions of our work are as follows:

1) We propose MTS-DVGAN, an unsupervised dual variational generative adversarial architecture for anomaly detection from multivariate time series to guarantee the security of industrial CPSs, which also compensates for the drawbacks of generating blurry samples and mode collapse of GAN to help enhance the stability.

2) To improve the detection rate of anomalies, we design an augmented module to widen the gap between reconstructed normal data and abnormal data by carrying out self-supervised learning on the embedding space. In detail, we impose contrastive constraints on the embedding process and reconstruction process by introducing another auxiliary VAE to obtain a higher quality feature center.

3) Three loss functions are designed to enhance the robustness and discriminative power of the proposed model, one of which is the feature center loss to enhance the stability of the generator, while the other two are augmented losses that extract essential features of input samples to obtain more discriminative embedding.

4) Experimental evaluations on three public datasets are conducted, including SWAT, WADI and NSL_KDD. Compared with six popular classical unsupervised methods, three GAN-based models (i.e., MAD-GAN and EGAN and GAN-AD), a VAE-based model (i.e., USAD) and a Transformer-based model (i.e. ADtrans), the results demonstrate that the proposed model MTS-DVGAN achieves the optimal performance, shows consistent improvement and stability.

## 2 RELATED WORK

Recently, deep generative models (such as GANs and VAEs) have become increasingly prevalent in the anomaly detection domain. Schlegl et al. [23] proposed AnoGAN, a deep convolutional GAN, to detect anomalies in medical images by learning a manifold of normal anatomical variability, accompanying a novel anomaly scoring scheme based on the mapping from image space to a latent space. Zenati et al. [19] constructed EGAN based on recently developed GAN methods with a learned encoder to achieve efficient anomaly detection. To handle the vulnerability of deep neural networks (DNN) to adversarial samples, Wang et al. [24] designed AC-GAN, an unsupervised attack detector on DNN classifiers for image classification based on class-conditional GANs. They calculated three detection statistics to model the clean data distribution using an auxiliary classifier. In contrast to the deep convolutional method, Li et al. [18] proposed GAN-AD utilizing long short-term memory recurrent neural networks (LSTM-RNN) to capture the hidden inherent correlations between multivariate time series produced by actuators and sensors. Li et al. [16] extended the GAN-AD model by utilizing the LSTM discriminator and generator networks, named MAD-GAN. This framework considers the entire variable set concurrently to capture the latent interactions among the variables and adopts a novel scoring schema called DR-Score, which combines both the discrimination and reconstruction loss to identify anomalies. Araujo-Filho et al. [25] constructed a fog-based intrusion detection system for CPSs, named FID-GAN, based on a GAN with a pretrained autoencoder, which detects attacks by calculating the reconstruction loss. They argued that introducing the pretrained encoder could accelerate the process of anomaly detection.

In addition, Su et al. [10] proposed OmniAnomaly, which combines VAE [26], [27] and GRU (a variant of RNN) for multivariate time series anomaly detection. This approach obtains the anomaly result by comparing the anomaly score of a data sample against a threshold selected via an automatic threshold selection algorithm. Unlike OmniAnomaly, Audibert et al. [7] designed an unsupervised anomaly detection model for Orange based on autoencoders using an adversarial training strategy, named USAD. They argued that the adoption of an adversarial training framework is beneficial to speed up training.

Our MTS-DVGAN differs from all the above models. We compare the structure of our model with all these models, as illustrated in Figure 3. In addition to the structural differences, more importantly, we take advantage of both feature center alignment and contrastive constraints to make the training process more stable and obtain a high-quality embedding space. For general understanding, we also analyze the strengths and weaknesses of each model as shown in Table 1.

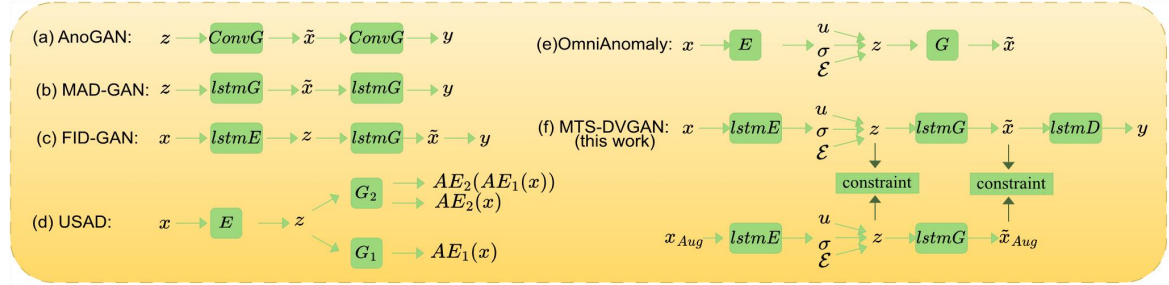

Figure 3: Illustration of the structure of (a) AnoGAN, (b) MAD-GAN, (c) FID-GAN, (d) USAD, (e) OmniAnomaly, (f) MTS-DVGAN, where E, G, and D are the encoder, generative and discriminative networks, respectively. $Conv*$ and $lstm*$ represent convolution and LSTM structures, respectively. x and x̃ are the input and generated samples (time series). z is the latent vector. μ and σ are the mean and variance of the Gaussian distribution that the input samples obey in latent space. ε is random noise. y is a binary output that denotes real or synthesized time series.

Additionally, Transformers [28] have shown great ability of modeling sequential data. Based on transformers, Xu et al [29] proposed ADtrans with an anomaly-attention mechanism to learn association discrepancy from time series for unsupervised detection of anomaly points. They identified anomalies by calculating anomaly scores which incorporated the learned association discrepancy into the reconstruction criterion. Although ADtrans [29] achieved good performance, its weakness is the complex and deep network structure resulting high training costs. Therefore, it may not be suited for resource-constrained industrial equipment.

Table 1. Strengths and weaknesses of each method (G, D and AE denote the generative, discriminative and autoencoder networks. Cell means the unit of network structure)

| Method | Component & cell | Strengths & Weaknesses | Applications |
|---|---|---|---|
| AnoGAN [23] | G, D & convolutions | **Strengths:** good extraction ability<br>**Weaknesses:** failed to handle the temporal dependence of multivariate time series; Failed to address the unstable gradient of G | Medical image |
| AC-GAN [24] | G, D & ResNet | **Strengths:** good extraction ability, could detect adversarial example<br>**Weaknesses:** high training cost and long training time, failed to address the unstable gradient of G | Image classification |
| EGAN [19] | AE, G, D & convolutions | **Strengths:** good extraction ability; fast test speed<br>**Weaknesses:** failed to handle the temporal dependence of multivariate time series; failed to address the unstable gradient of G | Image and network intrusion |
| MAD-GAN [16] | G, D & LSTM | **Strengths:** extract temporal dependencies of time series<br>**Weaknesses:** failed to address the unstable gradient of G; failed to identify the abnormal samples that distribute near normal samples | Multivariate time series |
| FID-GAN [25] | AE, G, D & LSTM | **Strengths:** extract temporal dependencies of time series; low delay<br>**Weaknesses:** failed to identify the abnormal samples that distribute near normal samples | Multivariate time series |
| USAD [7] | Two AE & linear | **Strengths:** fast, stable<br>**Weaknesses:** failed to extract the temporal dependencies; failed to identify the abnormal samples that distribute near normal samples | Multivariate time series |
| OmniAnomaly[10] | VAE & RNN | **Strengths:** modeling timing dependencies explicitly<br>**Weaknesses:** failed to identify the abnormal samples that distribute near normal samples | Multivariate time series |
| ADtrans [29] | Transformer | **Strengths:** good performance<br>**Weaknesses:** high training cost, high delay | Multivariate time series |
| MTS-DVGAN (ours) | Two VAE, G, D & LSTM | **Strengths:** extract temporal dependencies; alleviate the unstable gradient of G; identify the abnormal samples that distribute near normal samples; low delay; stable<br>**Weaknesses:** converge a little slower during training, within tolerable limits | Multivariate time series |

## 3 PRELIMINARY INFORMATION

### 3.1 LSTM

CNN-based and RNN-based models are two popular deep learning analysis models for feature extraction from multivariate time series [30]. In general, CNNs can capture short-term local features and perform better in extracting spatial dependencies from original data, while RNNs are more suitable to extract temporal dependencies from time series. However, traditional RNNs fail to capture long-term time dependence and are hard to train due to gradient explosion and gradient vanishing problems.

Long short-term memory network (LSTM) [31] was introduced to overcome these problems. As a variant of RNN, LSTM can recursively forget, retain and update information in recurrent neurons by adding gated units into RNN, namely forgetting gate $f_t$, input gate $i_t$ and output gate $o_t$, respectively. These three gate units allow LSTM to dynamically process the memory of the neural network, which can not only remember useful information in the long term of the time series but also forget low-value information, thus greatly improving the efficiency and quality of feature extraction.

### 3.2 VAE and GAN

As mentioned above, VAEs and GANs are two of the most popular generative models. Consisting of an encoder and a decoder, the VAE is not only a deep generative model but also a probabilistic graphical model first proposed by Diederik et al. [26] in 2013, which integrates the nonlinear feature representation capability of deep neural networks and the flexibility of probabilistic generation models. Given an observation $x$, the purpose of the VAE is to learn a model distribution $p(x)$ that is parameterized by the neural network so that the generated data can be as similar as possible to the real data distribution $q_{real}(x)$. However, because of the problems of the similarity measure function, VAE often synthesizes fuzzy samples [17].

On the other hand, the GAN [32] is another kind of generative model consisting of two adversarial networks: a generator and a discriminator. The generator tries to fool the discriminator by generating samples as similar as possible to the real ones, while the discriminator tries to distinguish as much as possible between the real and generated samples.

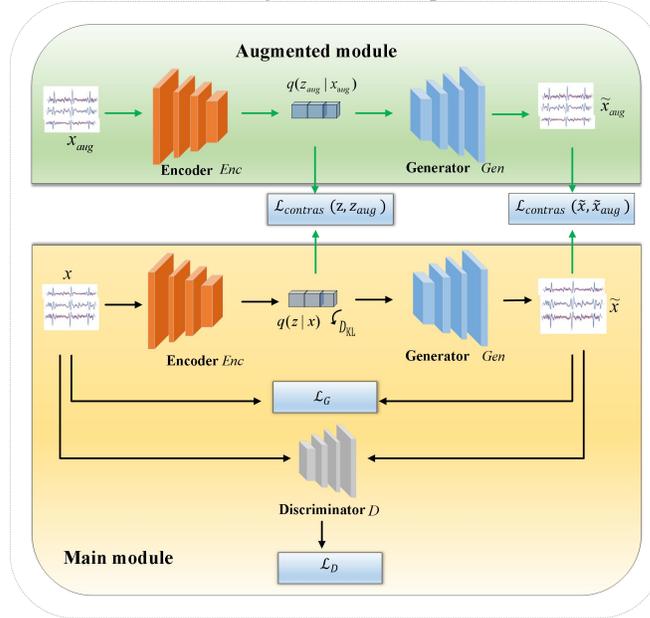

Figure 4: The architecture of the proposed model. It consists of two parts: the contrastive module (enclosed in a light green block) and the main module (enclosed in a light yellow block). The former aims to impose contrastive constraints on the latent space and the reconstruction space for generating high-quality samples, while the latter is for detecting anomalies. $Enc$, $Gen$ and $D$ are the encoder, generator and discriminator networks, respectively. The encoder and generator of the contrastive module share weights with those of the main module.

## 4 METHODOLOGY

In this section, we first formulate the problem statement of anomaly detection. Then, we describe the proposed architecture. Finally, we introduce two novel loss functions to enhance the performance and robustness of our model. The descriptions of the mathematical symbols used in this section are shown below:

Table 2. Description of the symbols used in Section 4

| Symbol | Description |
| --- | --- |
| $x_t$ | Multivariate time series at time step $t$ |
| $X$ | Observations/samples consists of a fixed number (window size) of $x_t$ |
| $p(z)$ | Prior distribution |
| $q_{real}(x)$ | Distribution of real data |
| $Enc$ | The encoder network |
| $Gen$ | The generator network |
| $D$ | The discriminator network |
| $q_\theta(z|x)$ | Posterior distribution |
| $z$ | Latent variable |
| $N(\mu, \sigma^2)$ | Gaussian distribution |
| $N(0, 1)$ | Standard Gaussian distribution |
| $\tilde{x}$ | Reconstructed data |
| $p_\phi(x|z)$ | Generative distribution of the $Gen$ |
| $f_{center}(x)$ | Feature center of $x$ |
| $fea_D$ | Features on an intermediate layer of the discriminator network $D$ |
| $Enc_l(x)$ | The hidden representation of the $l$-th layer of network $Enc$ |
| $x_{aug}$ | The augmented sample of $x$ |
| $\mathcal{L}_R$ | Reconstruction loss |
| $\mathcal{L}_D$ | Discrimination loss |

### 4.1 Problem statement

In our study, we focus on multivariate time series collected from industrial CPSs. The objective of anomaly detection is to assign binary (1 for abnormal and 0 for normal) labels to the collected time series $x_t$. As historical values are beneficial for understanding current data in time series modeling, we divide the multivariate time series into a set of multivariate subsequences using a sliding window technique [33]. Then, instead of just using independent $x_t$, a subsequence is used to calculate the anomaly result.

Therefore, we define these divided subsequence time series as $x = \{x_1, x_2, \cdots, x_k\} \in R^{N \times k}$, and $x_t = [x_t^1, x_t^2, ..., x_t^N]$ is a $N$-dimensional vector at time $t$ ($1 \leq t \leq k$), where $k$ is the sliding window size.

### 4.2 MTS-DVGAN architecture

Industrial CPSs continuously produce sensor and actuator data (in the form of multivariate time series), which can be used to monitor abnormal events during the system operation. In multivariate time series, we find that they are not independent of each other but exist long-term dependencies among them, which reflects the normal operation states of the CPS systems and are beneficial to detect anomalies. To exploit such valuable dependencies and model the normal patterns for anomaly detection, we propose a novel unsupervised dual variational generative adversarial model named MST-DVGAN which is able to learn the temporal dependencies by adopting deterministic hidden variables.

The architecture of the MTS-DVGAN model is shown in Figure 4, which consists of two parts: an augmented module that imposes contrastive constraints on the embedding space to enhance the discrimination ability of the proposed model and a main module that learns to detect anomalies from a large amount of MTS based on an anomaly score (See Section 5 for details). Specifically, the main module contains a variational inference network $Enc$ for learning a mapping: $F: x \rightarrow z$ that maps input samples into latent space; a generator $Gen$ for synthesizing $\tilde{x}$ based on noise randomly sampled from the latent space with an optimization objective $\mathcal{L}_G$; and a discriminator $D$ for distinguishing whether input samples are real or synthetic with $\mathcal{L}_D$ as its optimization objective. To improve the power of discriminating abnormal samples that are distributed near normal samples, we design an augmented module to capture more discriminative features, inspired by contrastive learning which has achieved promising results in computer vision field [34], [35]. The augmented module consists of an $Enc$ and a $Gen$ which are similar to the corresponding components in the main module and share parameters. The only difference between the two modules is that the augmented module handles augmented samples $x_{aug}$ rather than original samples $x$ (see Section 4.4 for more details). Then, a contrast constraint ($\mathcal{L}_{contras}(z, z_{aug})$) can be imposed on the latent space to extract more discriminative features from normal samples. Meanwhile, similar contrast constraint $\mathcal{L}_{contras}(\tilde{x}, \tilde{x}_{aug})$ is imposed on the reconstruction space to force synthesizing samples with more essential features. It is worth noting that all the encoder $Enc$, the generator $Gen$ and the discriminator $D$ of the both modules are constructed based on LSTM-RNNs.

**Main module:** As shown in the bottom half of Figure 4, the main module contains three components: the encoder network $Enc$, the generator network $Gen$, and the discriminator network $D$. We use $\theta$, $\phi$ and $\xi$ to represent their parameter sets.

We use an LSTM-based variational inference network $Enc$ to parameterize the variational posterior distribution $q_\theta(z|x)$ by mapping a multivariate time series $x$ with multiple time steps in a latent space. The posterior distribution $q_\theta(z|x)$ belongs exclusively to $x$ and stores the latent variable $z$. In general, $q_\theta(z|x)$ is always assumed to obey a general Gaussian distribution $N(\mu, \sigma^2)$ and is constrained to fit the standard Gaussian distribution $N(0, 1)$ by using the Kullback–Leibler (KL) divergence [36]. Then, using the reparameterization trick [26], random samples $z$ from the determined posterior Gaussian distribution and all $z$ form a latent $Z$-space [7].

The generative network $Gen$ attempts to generate reconstructed data $\tilde{x}$ as similar as possible to $x$ based on randomly sampled data from the latent distribution and the trained parameters of the generative distribution $p_\phi(x|z)$. Therefore, a formula of the optimal objective of these two networks based on the Stochastic Gradient Variational Bayes theory [37] is as follows:

$$\mathcal{L}_{VAE}(\theta, \phi) = - D_{KL}[q_\theta(z|x) \parallel p(z)] \qquad (1)$$

where $z = \mu + \sigma \otimes \epsilon$ and $\epsilon \sim N(0,1)$.

Subsequently, the reconstructed data are passed to the discriminative network $D$, which tries to distinguish whether the input data are real or not. It assigns probability $y = D(x) \in [0,1]$ that $x$ is a real data sample and probability $1 - y$ means $x$ is a fake sample generated by $Gen(z)$ with $z \sim p(z)$. These two networks (i.e., $Gen$ and $D$) are trained together in a two-player minimax game. The discriminator $D$ tries to minimize the probability of identifying the generating data as real data, while the generator $Gen$ tries to maximize that same probability [38]. Specifically, the network $Gen$ tries to maximize the loss function:

$$\mathcal{L}_G(\phi) = E_{z \sim p(z)}[\log D(Gen(z))] \qquad (2)$$

while the network $D$ tries to minimize

$$\mathcal{L}_D(\xi) = E_{x \sim p_{real}(x)}[\log D(x)] + E_{z \sim p(z)}[\log(1 - D(Gen(z)))] \qquad (3)$$

In our model, since the prior distribution $p(z)$ has been approximated by the encoder $Enc$ with the posterior distribution $q_\theta(z|x)$, which is closer to the true distribution of $x$ in the latent $Z$-space, the generator $Gen$ samples $z$ are forced from the learned posterior distribution instead of random noise (i.e., $p(z)$), enabling it to generate higher quality and adequately diverse samples for alleviating the instability issue. Thus, Eq. (2) and Eq. (3) can be modified as follows:

$$\mathcal{L}_G(\phi) = E_{z \sim q_\theta(z|x)}[\log D(Gen(z))] \tag{4}$$

$$\mathcal{L}_D(\xi) = E_{x \sim p_{real}(x)}[\log D(x)] + E_{z \sim q_\theta(z|x)}[\log(1 - D(Gen(z)))] \tag{5}$$

To further enhance the stability of $Gen$, we design a feature center loss objective instead of using cross-entropy loss, which is explained in Section 4.2 in detail.

**Augmented module:** To improve the discrimination ability of abnormal samples, the key point is to generate samples with unique characteristics to the corresponding original input. Thus, we construct an augmented module to obtain a compact high-quality normal data cluster in self-supervision mode through mutual guidance on the latent space, which could alleviate the failure of reconstruction-based methods to distinguish anomalies that lie near the normal data or that are distributed far from normal data but close to the latent dimension manifold of normal data in latent space. Since anomaly detection depends on the reconstruction error of our model (described in Section 5), it is important to separate abnormal data from normal data. Thus, we also impose the contrastive constraint on the reconstruction space to enlarge the gap between the reconstruction error of normal data and abnormal data.

As shown in the top half of Figure 4, the augmented module contains two components: an encoder and a generative network to impose contrastive learning on the latent space and reconstruction space (See Section 4.4 for details), respectively. It is worth noting that these two components share parameters with the corresponding networks in the main module.

### 4.3 Feature center loss

When training GAN, we have to manage the unstable gradient of the generator, which has been theoretically verified by recent works [3], [21], [22]. To address this issue, the key is to ensure that the $Gen$ can reconstruct higher-quality samples that are as similar as possible to the original ones. We believe that the feature centers of similar samples should also be similar, i.e., the distance between them should be small. Thus, we propose to use the distance between the feature centers of the generated samples and the real samples as the optimization objective for the generator $Gen$ instead of the original loss symmetry with the discriminator $D$. A smaller distance means that the feature center of the generated sample is closer to that of the real one. Therefore, in the training process, $Gen$ tries to maximize the following loss function:

$$L_{GC} = dis\,[f_{center}(x), f_{center}(G(Enc(x)))],$$

$$f_{center}(x) = \frac{1}{n}\sum_j^n fea_D(x),$$

$$f_{center}(G(E(x))) = \frac{1}{n}\sum_j^n fea_D(G(E(x))) \tag{6}$$

where $dis[,]$ denotes the distance function (here, Euclidean distance), $fea_D$ denotes features on an intermediate layer of the discriminator network $D$, and $f_{center}(x)$ and $f_{center}(G(E(x)))$ are the feature centers of input samples and reconstructed ones, respectively. For simplicity, the input of the last fully connected layer of discriminator $D$ is regarded as the feature $fea_D$ in our experiment. Notably, the feature center is calculated using mini-batch data in the training stage.

In addition, let $Enc_l(x)$ denote the hidden representation of the $l$-th layer of network $Enc$. Thus, we can replace the discriminator loss of Eq. (5) and the generator loss of Eq. (6) with the following Eq. (7) and Eq. (8) respectively.

$$\mathcal{L}_D = E[log D(x)] + E[1 - log D(G(Enc_l(x)))] \tag{7}$$

$$\mathcal{L}_{GC} = dis\,[f_{center}(x), f_{center}(G(Enc_l(x)))] \tag{8}$$

Although Eq. (6) looks similar to CycleGAN [39], there are three differences:

1) Difference in research targets: our target is to alleviate the issue of unstable gradient of the generator while CycleGAN's is to reduce the possible embedding space of the two mappings G and F.

2) Difference in methods: our Eq. (6) calculates the L2 norm between the feature center of generated samples and the feature center of real samples, while CycleGAN calculates the L1 distance between the cyclically generated samples and the original sample of the source domain.

3) Difference in sample generation: Our generator synthesizes samples based on the randomly sampling from the latent space generated by the encoder, while CycleGAN synthesizes target samples based on source domain samples and synthesizes source domain samples based on target samples.

### 4.4 Augmented loss

Since the characteristics of contrastive learning are that it can narrow similar samples and exclude dissimilar samples, we hope to expand the distance to the normal cluster of outliers or their latent dimension manifolds by generating compact intraclass and sparse interclass distributions. Additionally, training with perturbation data can also enable our model to be more robust.

Given an original sample $x$, $x_{aug}$ denotes the corresponding augmented sample that is transformed from the original data $x$ using the jitter-and-scale strategy [40]. Intuitively, $x$ and $x_{aug}$ belong to the same class, obey the same distribution in the latent space, and still belong to the same cluster after reconstruction. Then, to force a compact cluster in latent space, we minimize the augmented loss between $q(z|x)$ and $q(z_{aug}|x_{aug})$ during the training phase. Similarly, to improve the quality of reconstructed samples, the distance between $p(x|z)$ and $p(x_{aug}|z_{aug})$ is also optimized to synthesize samples with essential characteristics to improve the detection accuracy. The two augmented losses are formulated as follows:

$$\mathcal{L}_{contras}(z, z_{aug}) = ||z - z_{aug}||_2^2 \tag{9}$$

$$\mathcal{L}_{contras}(\tilde{x}, \tilde{x}_{aug}) = ||\tilde{x} - \tilde{x}_{aug}||_2^2 \tag{10}$$

Instead of using InfoNCE loss [34] like most contrast learning methods [41] [42], we adopt simple Euclidean distance as the augmented loss because 1) we find that our model can already achieve satisfactory performance and 2) resource-constrained industrial equipment (i.e., sensors) may not be able to support complex calculations.

Thus, according to Eq. (1), Eq. (8), Eq. (9), and Eq. (10), the total loss function of the $Enc$ and $Gen$ modules can be formalized as:

$$\mathcal{L}_{Enc}(\theta) = \mathcal{L}_G - D_{KL}[q_\theta(z|x) \| p_\theta(z)] - \alpha \mathcal{L}_{augment}(z, z_{aug}) \tag{11}$$

$$\mathcal{L}_{Gen}(\phi) = \mathcal{L}_G + \mathcal{L}_{GC} - \beta \mathcal{L}_{augment}(\tilde{x}, \tilde{x}_{aug}) \tag{12}$$

where $\alpha$ and $\beta$ are the hyperparameters to balance the weights of the corresponding augmented losses.

Therefore, we update the encoder $Enc$, the generator $Gen$ and the discriminator $D$ using Eq. (11), Eq. (12) and Eq. (7) during the training process, respectively. Using these losses for joint training of the three modules has the following three advantages: 1) Eq. (11) enables the model to learn essential sample features that can enhance the ability to discriminate abnormal samples; 2) since Eq. (7) increases the degree of distinction, it can alleviate the vanishing gradient problem; and 3) when the generated samples are good enough, the feature center loss is close to zero, making the training more stable.

In detail, the whole training procedure of the proposed unsupervised anomaly detection for multivariate time series data is described in Algorithm 1.

---
**ALGORITHM 1: Training Procedure of the Proposed MTS-DVGAN**

---

**Require:** $n$, the batch size. $\theta$, initial $Enc$ network parameters. $\phi$, initial $Gen$ network parameters. $\xi$, initial $D$ network parameters. $m$, the max iterative number. $step$, current iterative number.

**while** $step \leq m$ **do**

    $x_k \leftarrow$ Sample a batch from the real samples $p_{real}$

    $z \leftarrow Enc(x_k)$

    $\tilde{x}_g \leftarrow Gen(z)$

    $z_f \leftarrow$ Sample a batch from random noise $p_z$

    $x_f \leftarrow Gen(z_f)$

    $D_{KL} \leftarrow KL(q_\theta(z|x_k)||p_\theta(z))$

    $\mathcal{L}_D \leftarrow -(\log(1 - D(x_f)) + \log(1 - D(\tilde{x}_g)) + \log(D(x_k)))$

    $f_{center}(x_k) \leftarrow$ Calculate the feature center $\frac{1}{n}\sum_j^n fea_D(x_k)$ of $x_k$

    $f_{center}(x_g) \leftarrow$ Calculate the feature center $\frac{1}{n}\sum_j^n fea_D(x_g)$ of $\tilde{x}_g$

    $\mathcal{L}_G \leftarrow$ Calculate the similarity between $f_{center}(x_k)$ and $f_{center}(x_g)$ using Euclidean distance

    $x_{aug} \leftarrow$ Transform the original data $x_k$ to obtain its corresponding augmented sample

    $z_{aug} \leftarrow Enc(x_{aug})$

    $\tilde{x}_{aug} \leftarrow Gen(z_{aug})$

    $\mathcal{L}_{contras}(\tilde{x}, \tilde{x}_{aug}) \leftarrow ||z - z_{aug}||_2^2$

    $\mathcal{L}_{contras}(z, z_{aug}) \leftarrow ||\tilde{x}_g - \tilde{x}_{aug}||_2^2$

    $\xi \xleftarrow{+} \Delta_\xi(\mathcal{L}_D)$

    $\phi \xleftarrow{+} \Delta_\phi(\mathcal{L}_G + \mathcal{L}_{GC} - \beta \mathcal{L}_{augment}(\tilde{x}, \tilde{x}_{aug}))$

    $\theta \xleftarrow{+} \Delta_\theta(\mathcal{L}_G - D_{KL} - \alpha \mathcal{L}_{augment}(z, z_{aug}))$

    $step \leftarrow step + 1$

**end while**

---

## 5 ANOMALY DETECTION WITH MTS-DVGAN

Obviously, the discriminator $D$ learns to differentiate between real data and generated data. Moreover, the generator $Gen$ can also contribute to classification tasks, as shown by [16], [18], [23]. Therefore, we integrate the reconstruction loss $\mathcal{L}_R$ and discrimination loss $\mathcal{L}_D$ to compute an anomaly detection score for efficiently detecting anomalies. The $\mathcal{L}_R$ corresponds to the residual difference between a sample $x$ and its reconstruction $\tilde{x}$, i.e., the difference between $x$ and $Gen(z)$ when the representation $z$ of $x$ in the latent space is passed to the generator $Gen$, while the $\mathcal{L}_D$ corresponds to the discriminator's output, which implies whether $x$ is abnormal or not. The $\mathcal{L}_R$ measures how much a sample seems to be abnormal, as data that lie far away from the reconstructed ones by the generator are likely abnormal.

To calculate $\mathcal{L}_R$, the representation $z$ of the sample $x$ in the latent space is needed, i.e., the vector $z$ that represents the most similar data to $x$, to provide to the generator $Gen$ for reconstructing $\tilde{x}$. To this end, the encoder $Enc$ is introduced to continuously learn the distribution of $x$ in latent space. Consequently, the mapping from $x$ to $z$ (i.e., $Enc(x): x \to z$) and the mapping from $z$ to $x$ (i.e., $Gen(z): z \to x$) are explicitly established, which can enable the $Gen$ to generate samples that are as similar as possible to the original ones. Thus, the detection rate of the model is improved.

Since the reconstruction and discrimination losses, $\mathcal{L}_R$ and $\mathcal{L}_D$ respectively, measure how much a time series data sample seems to be the result of an anomaly event, we define an anomaly detection score $RD_{score}$ as a combination of these two losses as

$$RD_{score} = \lambda \mathcal{L}_D + (1-\lambda)\mathcal{L}_R \quad (13)$$

where $\lambda \in [0,1]$ is a hyperparameter that balances the contributions of $\mathcal{L}_R$ and $\mathcal{L}_D$ in the anomaly detection score. Note that if $\lambda = 1$, only the reconstruction loss is considered for computing $RD_{score}$. In the same way, if $\lambda = 0$, only the discrimination loss is considered. Note that to calculate the reconstruction loss $\mathcal{L}_R$, the mean square error (MSE) is employed as its indicator. Additionally, if one of the $\mathcal{L}_R$ and $\mathcal{L}_D$ indicates an attack, the signal will be identified as an attack, then the true positive rate will indeed approach 1. While this strategy effectively detects most abnormal behaviors, it may also lead to a corresponding increase in the false positive rate. This increase may be unacceptable for certain industrial systems where even a brief downtime can result in significant financial loss. For instance, the shutdown of an automobile manufacturing plant for just one minute may cost up to $20,000 as reported [43]. Therefore, we argue that the true positive rate and false positive rate should be balanced according to the specific requirements of individual industrial scenarios. Experimentally, the lambda is iterated from the interval [0, 1] with a step of 0.01, and the final lambda takes the value achieving the optimal result. Our detection procedure is shown in Algorithm 2.

---
**ALGORITHM 2: Our Anomaly Detection System**
---

**Require: X**, the evaluated dataset; $n$, the batch size; the trained networks $Enc$, $Gen$ and $D$ by Algorithm 1.

**for** each evaluated data sample $x$ in **X do**

    Calculate $z = Enc(x)$

    Calculate $\mathcal{L}_R(x) = \frac{1}{n}\sum_{i=1}^{n}(Gen(z_i) - x_i)^2$

    Calculate $\mathcal{L}_D(x)$

    Calculate $RD_{score} = \lambda \mathcal{L}_D + (1-\lambda)\mathcal{L}_R$

    Estimate whether $x$ is an anomaly or not using $RD_{score}$

**end for**

---

## 6  EXPERIMENTAL EVALUATION

This section describes the information of the evaluated experiments. We first introduce the details of the datasets used in our experiment. Then, we present the evaluation metrics and implementation details of our model.

### 6.1  Dataset Description

We evaluate the proposed MTS-DVGAN on two industrial process datasets, the water distribution (WADI) and the SWAT datasets, to verify the detection capability of the model on CPSs. Additionally, we also conduct an evaluation on the NSL_KDD dataset to show that our model can also achieve satisfactory performance on high-dimensional network activity data. Each of the three datasets is described below.

### 6.1.1 NSL_KDD dataset

The NSL_KDD dataset[1] [44] is refined from the KDD CUP99 dataset [45], which is a popular network traffic dataset prepared by Stolfo et al. [46] and is built based on the data collected in DARPA'98 intrusion detection evaluation testbed [47]. It contains a training set and a testing set. The former consists of approximately 125,973 records and is constructed from raw tcpdump connections from seven weeks of network traffic, while the latter has 22,544 records that are constructed from two weeks of network traffic. Each record in both sets contains 41 features, a label representing normal or not and a specific attack type. The dataset has 38 kinds of attacks, of which 24 are in the training set and 14 are in the test set, i.e., the data in the test set come from a probability distribution different from that of the training data, making the detection task more realistic. These 38 attacks can be divided into four categories in total: 1) surveillance and other probing attacks, e.g., port scanning; 2) unauthorized access from a remote machine (R2L) attacks, such as guessing password; 3) denial-of-service (DOS) attacks, e.g., syn flood; and 4) unauthorized access to local superuser with root privileges (U2R) attacks, e.g., various "buffer overflow" attacks. The details of the attacks in the training set are shown in Table 3.

Table 3. Details of the attacks in the NSL_KDD dataset

| Attack category | Specific attack types |
|---|---|
| DOS | Back, land, neptune, pod, smurf, teardrop |
| Probing | Ipsweep, nmap, portsweep, satan |
| R2L | ftp_write, guess_passwd, imap, multihop, phf, spy, warezclient, warezmaster |
| U2R | buffer_overflow, loadmodule, perl, rootkit |

### 6.1.2 SWAT dataset

The SWAT dataset [6] was built by the iTrust institution from a testbed of a real-world water treatment plant[2], which consists of a six-stage process of water treatment with several actuators and sensors, as shown in Figure 5. It consists of 495,000 normal records and 449,921 normal and attack records, in which each record contains a label and 51 attributes of actuator and sensor values. All of the data in the training set were collected over seven days of normal operation, while the testing set was recorded over four days in which four attack categories and a total of 36 different attack types were launched. The details of these attacks are summarized in Table 4.

Table 4. Details of the attack types in the SWAT dataset

| Attack type | Description | Number of attacks |
|---|---|---|
| SSSP | Attack focuses on only one point in a single stage | 26 |
| SSMP | Attack focuses on two or more attack points on one stage | 4 |
| MSSP | Attack is performed on single points on multiple stages | 2 |
| MSMP | Attack is performed on two or more stages | 4 |

---

[1] https://www.unb.ca/cic/datasets/nsl.html

[2] https://itrust.sutd.edu.sg/itrust-labs-home/itrust-labs_swat/

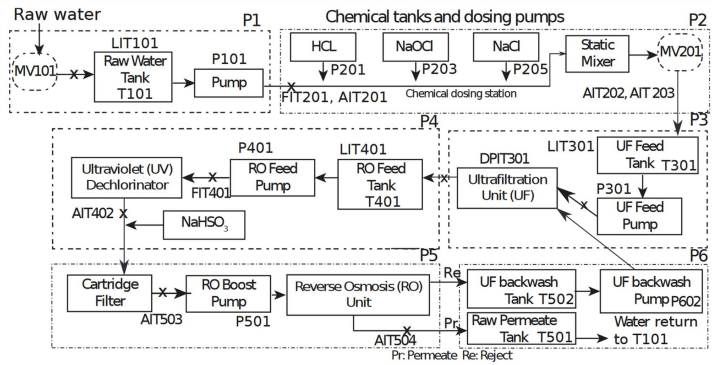

Figure 5: Overview of the SWAT testbed. P1~P6 represent its six different processes, each consisting of several sensors and actuators.

### 6.1.3 WADI dataset

The WADI dataset [3] is collected from the Water Distribution (WADI) testbed (shown in Figures 6 and 7), which is a natural extension of SWAT and was launched in 2016 by the Cyber Security Agency. The WADI testbed[4] not only carried out attacks and defenses on the PLCs and networks but also simulated the effects of physical attacks such as malicious chemical injections and water leakage, enabling researchers to verify the detection performance of the model against cyber and physical attacks. Its data collection process lasted for a total of sixteen days. During this period, the system operated normally during the first fourteen days, while during the last two days, cyber and physical attacks were conducted. Thus, it collected 1,048,571 normal and 172,801 normal and attack records, each with 126 features and a label.

We launched a testing, validation and training set for each dataset. The first two have both normal and attack records, and the latter has only normal data. The validation and training sets are used to obtain the algorithmic optimal hyperparameters, while the testing set is used to evaluate the detection performance of MTS-DVGAN.

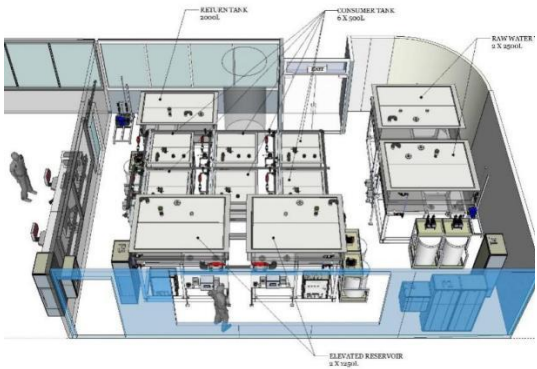
Figure 6. The diagram of WADI testbed.

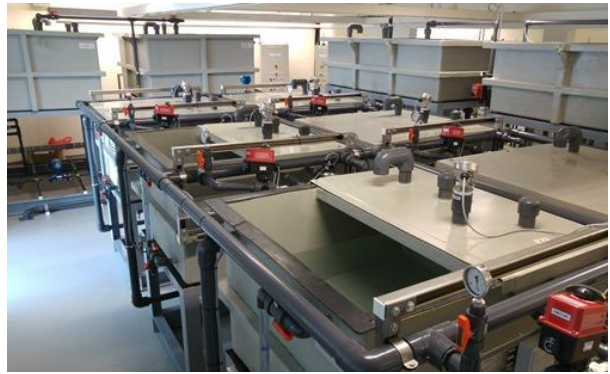
Figure 7. The layout of WADI testbed.

---

[3] https://itrust.sutd.edu.sg/itrust-labs_datasets/

[4] https://itrust.sutd.edu.sg/testbeds/water-distribution-wadi/

## 6.2 Data Preparation

### 6.2.1 Normalization

Before training, to eliminate the adverse effects caused by singular sample data and speed up convergence, all of the three datasets are normalized into [-1, 1] using the following formula:

$$x^{'} = 2 * \frac{x}{max(x)} - 1$$

### 6.2.2 Principal Component Analysis

To reduce the computation load, we use Principal Component Analysis (PCA) technique to filter out the informative features, by mapping the raw data into a low-dimensional principal space, instead of directly feeding high dimensional data to the MTS-DVGAN model.

To trade-off the model training cost and the anomaly detection performance, the principal components resolutions of the three datasets are selected: 12 (NSL_KDD), 5 (SWAT), and 8 (WADI).

## 6.3 Evaluation Metrics

To evaluate the detection performance of the proposed model, four standard metrics, namely, precision (a ratio of correct identification of positive samples to all positive samples), accuracy (a ratio of correct detection to all samples), recall (a ratio of correct detection of positive samples to all correct detection samples), and F1 score (a harmonic mean of recall and precision), are used:

$$Precision = \frac{TP}{FP + TP} * 100$$

$$Accuracy = \frac{TP + TN}{TP + FP + FN + TN} * 100$$

$$Recall = \frac{TP}{TP + FN} * 100$$

$$F1_{score} = 2 \times \frac{Precision \times Recall}{Precision + Recall} * 100$$

where FP, TP, FN, and TN indicate false positives, true positives, false negatives and true negatives, respectively. Furthermore, to fully verify the model, the area under the curve (AUC) of the receiver operating characteristic (ROC) curve is also used to visually evaluate the detection performance.

## 6.4 Implementation Details

Our objective in this work is to detect abnormal events from multivariate time series for industrial CPSs. Thus, adequately modeling the temporal dependency between data samples is quite important. For this purpose, we take a sliding window of size 30 to subdivide the original data streams into smaller time series with a shift length of 10, following Li et al. [16]. We use LSTM networks with a depth of 3 and 100 hidden units for the encoder, generator, and discriminator. Additionally, we adopt a latent space dimension of 15 in our study since we verified that the dimension of latent space with 15 could generate better samples. We extend the work of [16] by introducing two VAEs: one is for learning the distribution of data in latent space, and the other is for imposing a contrastive constraint on embedding space. Then, we conduct a comparative experiment with that work.

In our experiments, the proposed MTS-DVGAN is trained with the RMSProp optimizer for 500 epochs on each dataset, and the two hyperparameters are empirically set as $\alpha=0.1$ and $\beta=0.05$. Additionally, the weights of MTS-DVGAN are initialized with a truncated normal distribution. Meanwhile, the model is saved for each epoch. Then, we compute the $RD_{score}$ for all saved models and record their performances.

*6.4.1 Hyper-parameters*

The hyper-parameters of the MTS-DVGAN for each dataset are displayed in Table 5. For each dataset we set five parameters: 1) the learning rate for model training, 2) the size of windows corresponding to the length of the input sequential time series, 3) the dimension of embedding Z in the latent space, 4) the batch size for training, and finally 5) the signal number, which indicates the dimension of feature selection using principal component analysis (PCA) technique (details are listed at Section 6.2.2).

Table 5. Parameter details of the MTS-DVGAN on three datasets

| Datasets | Learning rate | Window size | Latent dimension | Batch size | Signal number |
|---|---|---|---|---|---|
| NSL_KDD | 1E-05 | 30 | 15 | 100 | 12 |
| SWAT | 5E-05 | 30 | 15 | 100 | 5 |
| WADI | 1E-04 | 30 | 15 | 1000 | 8 |

*6.4.2 Runtime Environment*

All experiments were conducted on a Genuine Intel(R) Xeon(R) Gold 6234 CPU @ 3.30 GHz with 125 G of RAM and an NVIDIA Quadro RTX 5000 with TensorFlow 2.5.0 environment.

The relevant packages and their versions used in our model implementation are listed as follows: python 3.8.5, tensorflow 2.5.0, cuda 11.0, scikit-learn 0.24.2, numpy 1.23.4, pandas 1.5.0.

## 6.5 Data and Code Acquisition

This paper uses three public datasets for experiments, i.e. SWAT, WADI, and NSL_KDD. The SWAT and WADI are collected by the iTrust institution and can be obtained from https://itrust.sutd.edu.sg/itrust-labs_overview/. The NSL_KDD dataset can be downloaded from https://www.unb.ca/cic/datasets/nsl.html. Source code of the project will be available at https://github.com/Platanus-hy/MTS-DVGAN.

## 7 ANALYSIS OF RESULTS

We evaluated the detection performance of MTS-DVGAN on the aforementioned three datasets NSL_KDD, WADI and SWAT. The subsequences were divided based on the sliding window strategy and then fed into the MTS-DVGAN model, as introduced earlier. For comparison, three GAN-based models (i.e., MAD-GAN [16], GAN-AD [18], and EGAN [19]) and a VAE-based model (i.e., USAD [7]) were applied to the three considered datasets. In addition, we compared with popular classic unsupervised anomaly detection methods including Principal Component Analysis (PCA), K-Nearest Neighbor (KNN), Feature Bagging (FB), Isolation Forest (IF), One Class Support Vector Machine (OCSVM) and Autoencoder (AE). A transformer-based method ADtrans[29] is also applied for comparison, which shows competitive performance on SWAT and WADI.

Table 6. Anomaly detection results of different methods on three datasets (%). We show the best performance by popular classic unsupervised methods (PCA, KNN, FB, IF, OCSVM and AE) with underlines, and indicate the overall optimal and suboptimal performance by (*) and (**) respectively.

| Datasets | Methods | Precision | Accuracy | Recall | F1 |
|---|---|---|---|---|---|
| SWAT | PCA | 24.92 | 42.67 | 21.63 | 22.96 |
| | KNN | 7.83 | 18.31 | 7.83 | 7.83 |
| | FB | 10.17 | 20.12 | 10.17 | 10.17 |
| | IF | 49.29 | 38.93 | 44.95 | 47.02 |
| | OCSVM | 45.39 | 43.64 | 49.22 | 47.23 |
| | AE | <u>72.63</u> | <u>65.21</u> | <u>52.63</u> | <u>61.43</u> |
| | EGAN | 40.57 | 71.86 | 67.73 (*) | 50.68 |
| | GAN-AD[5] | 93.33 | 94.80 (**) | 63.64 | 75.48 |
| | MAD-GAN | 17.27 | 73.05 | 30.95 | 22.11 |
| | USAD | 94.63 (**) | 89.87 | 59.65 | 73.17 |
| | ADtrans | 94.58 | 93.57 | 67.73 (*) | 78.93 (**) |
| | MTS-DVGAN | 99.0 (*) | 94.84 (*) | 66.93 (**) | 79.87 (*) |
| WADI | PCA | <u>39.53</u> | <u>45.29</u> | 5.63 | 9.86 |
| | KNN | 7.76 | 19.72 | 7.75 | 7.75 |
| | FB | 8.60 | 20.56 | 8.60 | 8.60 |
| | IF | 15.55 | 36.18 | 19.88 | 17.45 |
| | OCSVM | 13.98 | 30.89 | 20.62 | 16.66 |
| | AE | 34.35 | 38.27 | <u>34.35</u> | <u>34.35</u> |
| | EGAN | 11.33 | 46.73 | 37.84 | 17.43 |
| | GAN-AD[5] | 95.31 | 59.42 | 60.36 | 73.91 |
| | MAD-GAN | 94.29 | 86.10 | 90.83 | 92.52 |
| | USAD | 96.09 (**) | 94.81 (*) | 15.23 | 26.29 |
| | ADtrans | 93.46 | 91.78 | 98.24 (**) | 95.79 (**) |
| | MTS-DVGAN | 96.23 (*) | 94.63 (**) | 99.51 (*) | 97.84 (*) |
| NSL_KDD | PCA | 61.76 | <u>68.25</u> | 35.93 | 45.43 |
| | KNN | 46.21 | 53.98 | 17.38 | 25.26 |
| | FB | 49.99 | 57.21 | 20.46 | 29.04 |
| | IF | 56.74 | 62.27 | <u>55.19</u> | <u>55.95</u> |
| | OCSVM | 57.68 | 61.56 | 53.24 | 55.37 |
| | AE | <u>80.59</u> | 63.54 | 42.36 | 55.48 |
| | EGAN | 92.00 (*) | 78.64 | 95.82 (**) | 93.87 (*) |
| | GAN-AD[5] | 85.53 | 54.24 | 54.31 | 66.43 |
| | MAD-GAN | 83.66 | 64.83 | 70.53 | 76.53 |
| | USAD | 84.00 | 69.78 | 74.83 | 79.15 |
| | ADtrans | 90.20 | 81.29 (**) | 93.72 | 91.93 |
| | MTS-DVGAN | 91.17 (**) | 82.46 (*) | 96.64 (*) | 93.78 (**) |

## 7.1 Performance Comparison

Table 6 shows the performance of MTS-DVGAN and the eleven benchmark models mentioned above for comparison. The results of PCA, KNN, FB, AE and EGAN on the three datasets were obtained from [16], the missing Accuracy results were obtained through experimental testing. The results of IF and OCSVM on SWAT were obtained from [29], and the results of GAN-AD[5] on SWAT were obtained from [18] (the first five principal components are chosen). Other experimental results come from code implementation. For fair comparison, we also chosed the highest F1 score as results of the four popular models, as in [16], since the F1 score balances the recall and precision metrics.

From Table 6, the following can be observed:

1) For the SWAT dataset, in general, MTS-DVGAN achieves the optimal performance compared with the other eleven benchmark models. Compared with ADtrans (the suboptimal model), the MTS-DVGAN proposed in this paper is about 2.21% higher than ADtrans in average on three metrics except recall, and only slightly lower than ADtrans and EGAN on recall. In addition, on F1 score, the performance of our model is significantly better than that of classical models and other unsupervised comparison models including GAN-AD[5], MAD-GAN and USAD.

2) For the WADI dataset, compared with ADtrans (the suboptimal model), the model MTS-DVGAN proposed in this paper exceeds ADtrans (by 2.24% in average) on all four evaluation metrics. The recall and F1 score of the MTS-DVGAN are significantly higher by a large margin (84.28% and 71.55%) than those of the USAD, except for accuracy, which is slightly lower (0.18%) than USAD model, the performances on the other metrics are significantly better than that of classical models and the benchmark models.

3) For the NSL_KDD dataset, the accuracy and recall of MTS-DVGAN are higher than that of EGAN (the suboptimal model), although the latter achieves a slightly better F1 score and precision. Moreover, the performance of the former significantly outperformed the latter on both the WADI and SWAT datasets. In addition, our model performs much better than other competitive models, including ADtrans, on all four metrics. It is important to note that the NSL_KDD is a dataset of network traffic, but the SWAT and WADI datasets are collections of sensor values. According to Li et al. [16], EGAN is more suitable for dealing with network traffic data, which may be the reason for the above results that EGAN achieves a slightly better F1 score and precision on the NSL_KDD dataset.

Additionally, it is obvious that the results of MTS-DVGAN on four evaluation metrics are higher than those of MAD-GAN and GAN-AD for all datasets. Overall, the MTS-DVGAN achieves the optimal performance on the three datasets compared with the other the other six classical models and five GAN-based or Transformer-based detection models, which implies that the improvement of our model is indeed effective in enhancing detection performance.

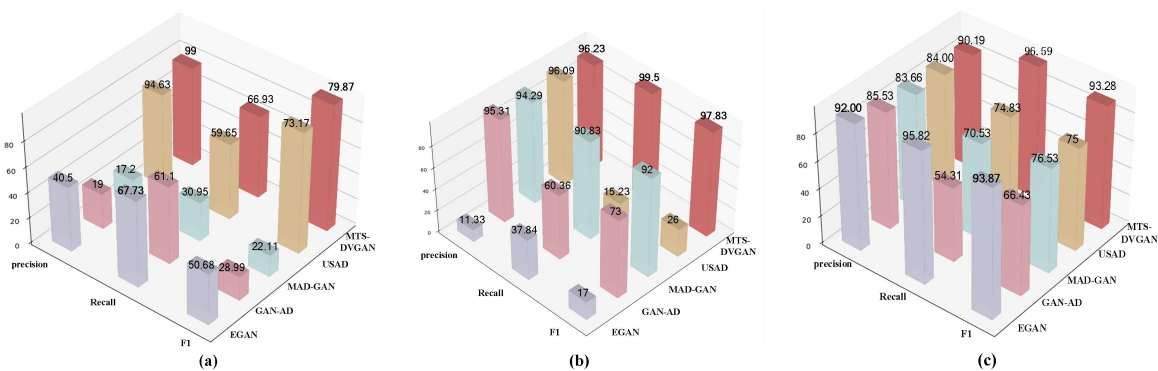

Figure 8: Histogram of performance for different models. (a) SWAT dataset. (b) WADI dataset. (c) NSL_KDD dataset. The performance metrics of different models are plotted in different colors. The numbers on top of the cuboid represent corresponding performance metrics.

For intuitive analysis, the performance of the GAN-based and VAE-based models on the three datasets are shown in Figure 8. This histogram illustrates the highest metric values attained by all models for each dataset, where the maximum values are displayed in bold format. It further shows that MTS-DVGAN outperforms its competitors overall.

### 7.2 Statistical Comparisons based on Friedman and Nemenyi test

For further comparison, based on the Friedman test [48] and Nemenyi test [49], we conduct statistical tests for comparisons of all the methods in Table 6 on the three datasets. According to the Friedman test, we rank each method for each dataset separately as shown in Table 7. Then, to explain and visualize the statistical test results, we calculate the critical difference (CD) between the average rank values (significance a=0.05) based on the Nemenyi test by the following equation:

$$CD = q_\alpha \sqrt{\frac{k(k+1)}{6N}}$$

where $q_\alpha$ is critical values, $\alpha$ is the significance level, $k$ denotes the number of methods and $N$ represents the number of datasets. In our statistical test, we set $\alpha = 0.05$, $k = 6$ and $N = 3$.

Then, we draw the Friedman test plot as shown in Figure 9. It is worth noting that the $x$-axis denotes rank values and the $y$-axis represents the different models. No overlap between two lines means there exists a significant difference between the corresponding two models.

As shown in Figure 9, there is overlap between almost arbitrary two lines, indicating no significant difference between these models. However, compared with all the comparison models, it can be observed that MTS-DVGAN has the higher rank values which indicates its superiority. In addition to the higher rank values, compared with ADtrans, we argue that MTS-DVGAN is still comparable, due to the deep and complex network structure of ADtrans which lead to high training costs and slow testing speed, which may be unacceptable for industrial systems that have requirements on latency. Therefore, in terms of training cost, detection speed and detection performance, our model is still superiority.

Table 7. Rank values of each comparative model on the three datasets

|  | MTS-DVGAN | ADtrans | USAD | MAD-GAN | GAN-AD | EGAN | AE | OCSVM | IF | FB | KNN | PCA |
|---|---|---|---|---|---|---|---|---|---|---|---|---|
| SWAT | 1 | 2 | 4 | 10 | 3 | 6 | 5 | 7 | 8 | 11 | 12 | 9 |
| WADI | 1 | 2 | 6 | 3 | 4 | 8 | 5 | 9 | 7 | 11 | 12 | 10 |
| NSL_KDD | 2 | 3 | 4 | 5 | 6 | 1 | 8 | 9 | 7 | 11 | 12 | 10 |
| Average | 1.33 | 2.33 | 4.67 | 6 | 4.33 | 5 | 6 | 8.33 | 7.33 | 11 | 12 | 9.67 |

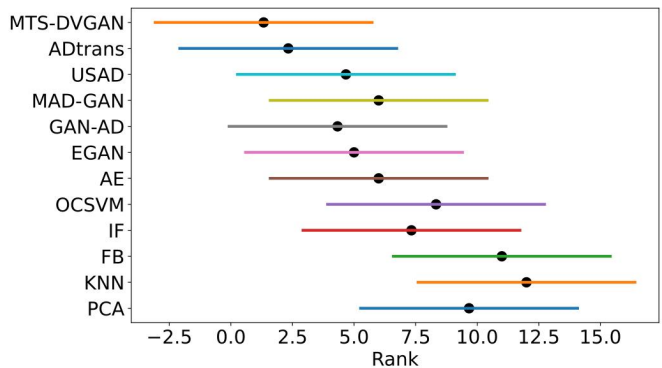

Figure 9: The Friedman test plot

### 7.3 Stability Analysis across Multiple Iteration Epochs

To evaluate the stability of our model during the training process, we draw the recall rate of each epoch on the three datasets and draw the recall rate of MTS-DVAN and the other three GAN-based methods (MAD-GAN, GAN-AD and EGAN) for comparison, as shown in Figure 10. The recall values of MTS-DVGAN on each dataset are significantly higher (nearly 30%) than those of the other three methods. On closer scrutiny, for the SWAT, WADI and NSL_KDD datasets, although the performance of MTS-DVGAN fluctuated, its amplitudes (nearly 10%, 4%, and 4%) were much smaller than those of MAD-GAN (nearly 20%, 11% and 22%) which seems to be the least oscillatory of the three compared models. These results demonstrate that the improvement measures (i.e., introducing an unsupervised dual variational generative adversarial model and the proposed feature center loss) can indeed strengthen the stability of our model effectively. Note that since the analysis is about the generator stability of GANs, we only compare with the GAN-based models here.

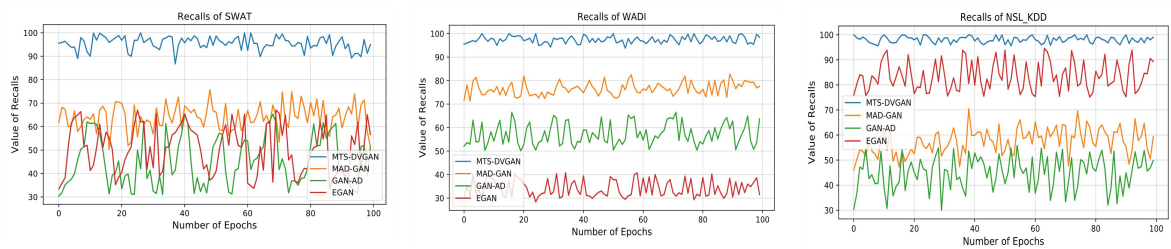

Figure 10: Values of recall as a function of the iteration epochs on the three datasets.

### 7.4 ROC Curve Analysis

Additionally, to study different contributions for the reconstruction and discriminant losses, we also draw the ROC curve and compare our results with FID-GAN [25], MAD-GAN [16] and GAN-AD [18], but not USAD as it only contains the reconstruction loss. A higher AUC means better detection results. Figure 11 depicts the ROC curves obtained by our model for the three public datasets. Similarly, Figure 12, Figure 13 and Figure 14 exhibit the ROC curves of the FID-GAN, MAD-GAN and GAN-AD models, respectively.

It is observed that the AUC value of our model is higher than the other three comparative models on the three datasets, demonstrating the superiority of the proposed model. The reconstruction loss $L_R$ and discrimination loss $L_D$ achieve comparable AUCs on different datasets, proving that both can be used to identify anomalies. Moreover, the AUC increases when combining the two losses, proving that the combination can integrate the advantages of both to achieve higher detection performance.

On the other hand, as shown in Figures 11-14, since the performance of the reconstruction loss of the other three GAN-based comparative models is lower than their discriminator loss, it may indicate that the discriminator is superior to the generator in identifying anomalies, which implies that our proposed improvements (such as the contrastive constraint imposed in the embedding space) could enhance the detection capability of the generator.

In addition, compared with GAN-AD (CNN-based), MAD-GAN (LSTM-based) significantly improves the performance, indicating that there are indeed complex temporal dependencies in multivariate time series, and modeling them can help with abnormal behavior since LSTM are better at learning temporal dependencies than CNN.

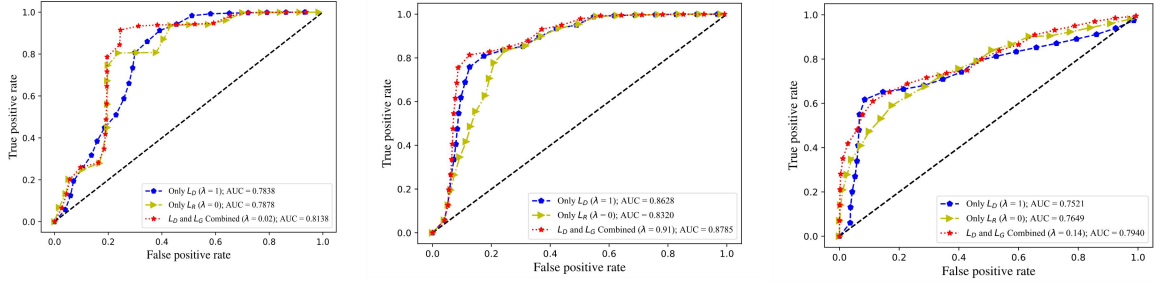

Figure 11: ROC curves of proposed MTS-DVGAN. (1) NSL_KDD dataset. (2) SWAT dataset. (3) WADI dataset.

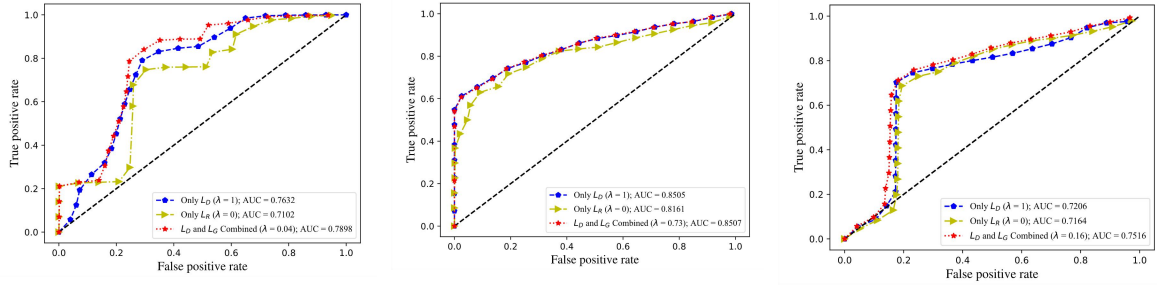

Figure 12: ROC curves of FID-GAN. (1) NSL_KDD dataset. (2) SWAT dataset. (3) WADI dataset.

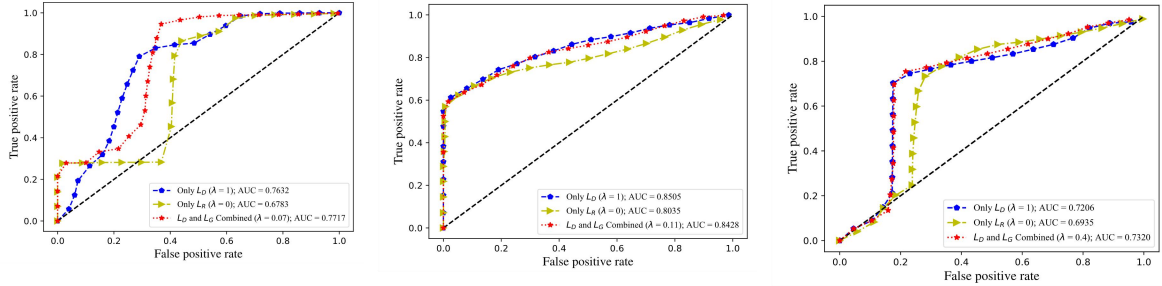

Figure 13: ROC curves of MAD-GAN. (1) NSL_KDD dataset. (2) SWAT dataset. (3) WADI dataset.

### 7.5 Ablation study

To evaluate the effects of each component in our architecture, we conduct an ablation study on the NSL_KDD, SWAT and WADI datasets by excluding each submodule from the MTS-DVGAN model. The ablation results are summarized in Table 8. We test three scenarios: (1) removing the contrastive module; (2) removing the encoder; and (3) using traditional binary cross entropy instead of the proposed feature center loss. The results agree with our assumptions that introducing the embedding space (by jointly training an encoder) can improve the detection performance. Furthermore, we find that

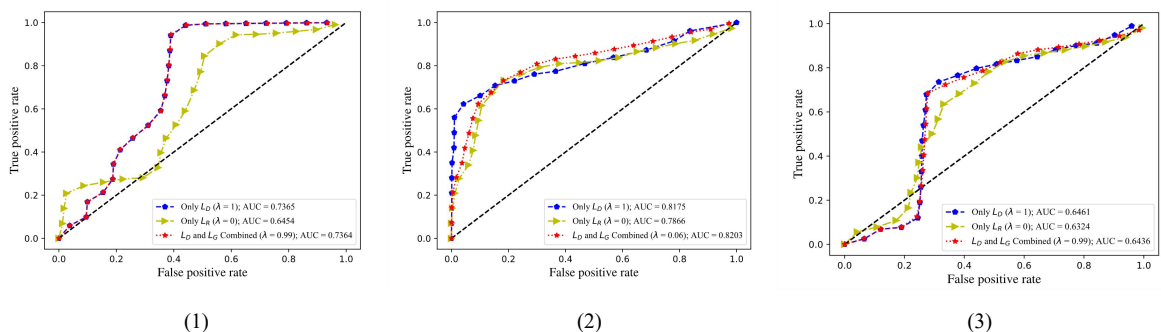

(1)                                 (2)                                 (3)

Figure 14: ROC curves of GAN-AD. (1) NSL_KDD dataset. (2) SWAT dataset. (3) WADI dataset.

removing the contrastive module leads to performance degradation, probably because the contrastive module can enhance the quality of the embedding space features. In other words, a higher quality embedding space yields higher performance. In addition, it is observed that the performance is worse without feature center loss, indicating its effectiveness for model refinement.

Table 8. Ablation study of the proposed MTS-DVGAN on the three datasets (F1 score (%))

| Settings | SWAT | WADI | NSL_KDD |
| --- | --- | --- | --- |
| MTS-DVGAN | 79.87 | 97.84 | 93.28 |
| wo. contrastive constraint | 77.68 (-2.19) | 95.61 (-2.23) | 91.40 (-1.88) |
| wo. encoder | 76.86 (-3.01) | 96.55 (-1.29) | 90.52 (-2.76) |
| wo. feature center loss | 78.98 (-0.89) | 96.89 (-0.95) | 91.90 (-1.38) |

### 7.6 Adaptive Attack Discussion

In this section, we discuss the adaptive attack assumption, i.e. assume that there are malicious adaptive attacks that have access to not only the Cyber-physical systems but also the detection method. If an attacker has access to the model, it may learn the features extracted by the model through some simulation approach and in turn produce samples that have similar characteristics, to bypass the anomaly detection model. But this behavior of the attacker will certainly increase its work factor.

From the viewpoint of features, the proposed model focuses on extracting the temporal dependencies and patterns from the sensor values generated by the Cyber-physical systems under normal working conditions. Therefore, to bypass the detection of the model, the attacker has to first continuously spend time and effort to learn the above dependencies and patterns. Then, it has to create attack behaviors that conform to the above dependencies and patterns. Only when these two steps are completed is it possible to bypass the detection of the model.

Under such an assumption, the attacker needs to assume a relatively strict prior knowledge, which is difficult to achieve in practice. Even if the requirements are met, it will take a lot of extra work to simulate a normal behavior to bypass detection, resulting in more time and calculation costs.

## 8 CONCLUSION AND FUTURE WORK

In this paper, we propose a novel unsupervised learning model, named MTS-DVGAN, to perform anomaly detection from multivariate time series generated in industrial CPSs. To identify such anomalies that lie near the normal samples in distribution or lie near the latent dimension manifold of the normal data cluster in embedding space, we design a contrastive module and two augmented losses to enlarge the reconstruction error gaps between abnormal and normal data by applying the contrastive constraint on the embedded space. To improve the detection rate, we explicitly model the data distribution in latent space by designing an encoder to force the generator to produce high-quality samples. Moreover, we design a feature center loss to enhance the stability of the generator.

For evaluation, we conduct extensive experiments on two real-world industrial datasets and a network traffic dataset. Experimental results show that the proposed approach can substantially improve the performance of anomaly detection. In addition, the comparative results indicate the superiority (both in performance and stability) of our MTS-DVGAN. In future research, we will investigate how to capture the evolution patterns of local and global status over time of industrial CPSs for anomaly detection and health diagnoses.

## ACKNOWLEDGMENTS

This paper is supported by the National Key Research and Development Program of China (No. 2022YFB3103402) and the National Natural Science Foundation of China (No. 62072200, No. 62172176, No. 62127808).

## REFERENCES

[1] H. Lasi, P. Fettke, H.-G. Kemper, T. Feld, and M. Hoffmann, "Industry 4.0," *Bus. Inf. Syst. Eng.*, vol. 6, no. 4, pp. 239–242, Aug. 2014, doi: 10.1007/s12599-014-0334-4.

[2] A. Humayed, J. Lin, F. Li, and B. Luo, "Cyber-Physical Systems Security—A Survey," *IEEE Internet Things J.*, vol. 4, no. 6, pp. 1802–1831, 2017, doi: 10.1109/JIOT.2017.2703172.

[3] J. Goh, S. Adepu, M. Tan, and Z. S. Lee, "Anomaly Detection in Cyber Physical Systems Using Recurrent Neural Networks," in *2017 IEEE 18th International Symposium on High Assurance Systems Engineering (HASE)*, Jan. 2017, pp. 140–145. doi: 10.1109/HASE.2017.36.

[4] R. Alguliyev, Y. Imamverdiyev, and L. Sukhostat, "Cyber-physical systems and their security issues," *Comput. Ind.*, vol. 100, pp. 212–223, Sep. 2018, doi: 10.1016/j.compind.2018.04.017.

[5] D. Kwon, H. Kim, J. Kim, S. C. Suh, I. Kim, and K. J. Kim, "A survey of deep learning-based network anomaly detection," *Clust. Comput.*, vol. 22, no. 1, pp. 949–961, Jan. 2019, doi: 10.1007/s10586-017-1117-8.

[6] J. Goh, S. Adepu, K. N. Junejo, and A. Mathur, "A Dataset to Support Research in the Design of Secure Water Treatment Systems," in *Critical Information Infrastructures Security*, G. Havarneanu, R. Setola, H. Nassopoulos, and S. Wolthusen, Eds., in Lecture Notes in Computer Science. Cham: Springer International Publishing, 2017, pp. 88–99. doi: 10.1007/978-3-319-71368-7_8.

[7] J. Audibert, P. Michiardi, F. Guyard, S. Marti, and M. A. Zuluaga, "USAD: UnSupervised Anomaly Detection on Multivariate Time Series," in *Proceedings of the 26th ACM SIGKDD International Conference on Knowledge Discovery & Data Mining*, in KDD '20. New York, NY, USA: Association for Computing Machinery, Aug. 2020, pp. 3395–3404. doi: 10.1145/3394486.3403392.

[8] H. Xu *et al.*, "Unsupervised Anomaly Detection via Variational Auto-Encoder for Seasonal KPIs in Web Applications," in *Proceedings of the 2018 World Wide Web Conference on World Wide Web - WWW '18*, 2018, pp. 187–196. doi: 10.1145/3178876.3185996.

[9] H. Ren *et al.*, "Time-Series Anomaly Detection Service at Microsoft," in *Proceedings of the 25th ACM SIGKDD International Conference on Knowledge Discovery & Data Mining*, Anchorage AK USA: ACM, Jul. 2019, pp. 3009–3017. doi: 10.1145/3292500.3330680.

[10] Y. Su, Y. Zhao, C. Niu, R. Liu, W. Sun, and D. Pei, "Robust Anomaly Detection for Multivariate Time Series through Stochastic Recurrent Neural Network," in *Proceedings of the 25th ACM SIGKDD International Conference on Knowledge Discovery & Data Mining*, in KDD '19. New York, NY, USA: Association for Computing Machinery, Jul. 2019, pp. 2828–2837. doi: 10.1145/3292500.3330672.

[11] B. Sun, P. B. Luh, Q.-S. Jia, Z. O'Neill, and F. Song, "Building Energy Doctors: An SPC and Kalman Filter-Based Method for System-Level Fault Detection in HVAC Systems," *IEEE Trans. Autom. Sci. Eng.*, vol. 11, no. 1, pp. 215–229, Jan. 2014, doi: 10.1109/TASE.2012.2226155.

[12] H. Choi, M. Kim, G. Lee, and W. Kim, "Unsupervised learning approach for network intrusion detection system using autoencoders," *J. Supercomput.*, vol. 75, no. 9, pp. 5597–5621, Sep. 2019, doi: 10.1007/s11227-019-02805-w.

[13] B. Sun, P. B. Luh, Z. O'Neill, and F. Song, "Building energy doctors: SPC and Kalman filter-based fault detection," in *2011 IEEE International Conference on Automation Science and Engineering*, Aug. 2011, pp. 333–340. doi: 10.1109/CASE.2011.6042429.

[14] H. Zenati, M. Romain, C.-S. Foo, B. Lecouat, and V. Chandrasekhar, "Adversarially Learned Anomaly Detection," in *2018 IEEE International Conference on Data Mining (ICDM)*, Nov. 2018, pp. 727–736. doi: 10.1109/ICDM.2018.00088.

[15] S. Li and J. Wen, "A model-based fault detection and diagnostic methodology based on PCA method and wavelet transform," *Energy Build.*, vol. 68, pp. 63–71, Jan. 2014, doi: 10.1016/j.enbuild.2013.08.044.

[16] D. Li, D. Chen, L. Shi, B. Jin, J. Goh, and S.-K. Ng, "MAD-GAN: Multivariate Anomaly Detection for Time Series Data with Generative Adversarial Networks." arXiv, Jan. 15, 2019. doi: 10.48550/arXiv.1901.04997.

[17] F. Harrou, M. N. Nounou, H. N. Nounou, and M. Madakyaru, "PLS-based EWMA fault detection strategy for process monitoring," *J. Loss Prev. Process Ind.*, vol. 36, pp. 108–119, Jul. 2015, doi: 10.1016/j.jlp.2015.05.017.

[18] D. Li, D. Chen, J. Goh, and S. Ng, "Anomaly Detection with Generative Adversarial Networks for Multivariate Time Series," arXiv.org. Accessed: Feb. 10, 2023. [Online]. Available: https://arxiv.org/abs/1809.04758v3

[19] H. Zenati, C. S. Foo, B. Lecouat, G. Manek, and V. R. Chandrasekhar, "Efficient GAN-Based Anomaly Detection." arXiv, May 01, 2019. doi: 10.48550/arXiv.1802.06222.

[20] T. Denouden, R. Salay, K. Czarnecki, V. Abdelzad, B. Phan, and S. Vernekar, "Improving Reconstruction Autoencoder Out-of-distribution Detection with Mahalanobis Distance." arXiv, Dec. 06, 2018. doi: 10.48550/arXiv.1812.02765.

[21] M. Arjovsky and L. Bottou, "Towards Principled Methods for Training Generative Adversarial Networks," presented at the International Conference on Learning Representations, Jul. 2022. Accessed: Feb. 02, 2023. [Online]. Available: https://openreview.net/forum?id=Hk4_qw5xe

[22] T. Salimans, I. Goodfellow, W. Zaremba, V. Cheung, A. Radford, and X. Chen, "Improved Techniques for Training GANs." arXiv, Jun. 10, 2016. doi: 10.48550/arXiv.1606.03498.

[23] T. Schlegl, P. Seeböck, S. M. Waldstein, U. Schmidt-Erfurth, and G. Langs, "Unsupervised Anomaly Detection with Generative Adversarial Networks to Guide Marker Discovery." arXiv, Mar. 17, 2017. doi: 10.48550/arXiv.1703.05921.

[24] H. Wang, D. J. Miller, and G. Kesidis, "Anomaly detection of adversarial examples using class-conditional generative adversarial networks," *Comput. Secur.*, vol. 124, p. 102956, Jan. 2023, doi: 10.1016/j.cose.2022.102956.

[25] P. Freitas de Araujo-Filho, G. Kaddoum, D. R. Campelo, A. Gondim Santos, D. Macêdo, and C. Zanchettin, "Intrusion Detection for Cyber–Physical Systems Using Generative Adversarial Networks in Fog Environment," *IEEE Internet Things J.*, vol. 8, no. 8, pp. 6247–6256, Apr. 2021, doi: 10.1109/JIOT.2020.3024800.

[26] D. P. Kingma and M. Welling, "Auto-Encoding Variational Bayes." arXiv, Dec. 10, 2022. doi: 10.48550/arXiv.1312.6114.

[27] D. J. Rezende, S. Mohamed, and D. Wierstra, "Stochastic Backpropagation and Approximate Inference in Deep Generative Models." arXiv, May 30, 2014. doi: 10.48550/arXiv.1401.4082.

[28] A. Vaswani *et al.*, "Attention is all you need," in *Proceedings of the 31st International Conference on Neural Information Processing Systems*, in NIPS'17. Red Hook, NY, USA: Curran Associates Inc., Dec. 2017, pp. 6000–6010. doi:/10.48550/arXiv.1706.03762.

[29] J. Xu, H. Wu, J. Wang, and M. Long, "Anomaly Transformer: Time Series Anomaly Detection with Association Discrepancy." arXiv, Jun. 29, 2022. doi: 10.48550/arXiv.2110.02642.

[30] J. L. Elman, "Finding Structure in Time," *Cogn. Sci.*, vol. 14, no. 2, pp. 179–211, 1990, doi: 10.1207/s15516709cog1402_1.

[31] S. Hochreiter and J. Schmidhuber, "LSTM can Solve Hard Long Time Lag Problems," in *Advances in Neural Information Processing Systems*, MIT Press, 1996. Accessed: Feb. 08, 2023. [Online]. Available: https://proceedings.neurips.cc/paper/1996/hash/a4d2f0d23dcc84ce983ff9157f8b7f88-Abstract.html

[32] I. J. Goodfellow *et al.*, "Generative Adversarial Networks." arXiv, Jun. 10, 2014. doi: 10.48550/arXiv.1406.2661.

[33] T. Sejnowski and C. R. Rosenberg, "Parallel Networks that Learn to Pronounce English Text," *Complex Syst*, 1987, Accessed: Jan. 18, 2023. [Online]. Available: https://www.semanticscholar.org/paper/Parallel-Networks-that-Learn-to-Pronounce-English-Sejnowski-Rosenberg/de996c32045df6f7b404dda2a753b6a9becf3c08

[34] A. van den Oord, Y. Li, and O. Vinyals, "Representation Learning with Contrastive Predictive Coding," Jul. 2018, doi: 10.48550/arXiv.1807.03748.

[35] O. J. Hénaff *et al.*, "Data-efficient image recognition with contrastive predictive coding," in *Proceedings of the 37th International Conference on Machine Learning*, in ICML'20, vol. 119. JMLR.org, Jul. 2020, pp. 4182–4192.

[36] S. Kullback and R. A. Leibler, "On Information and Sufficiency," *Ann. Math. Stat.*, vol. 22, no. 1, pp. 79–86, 1951, doi: 10.1214/aoms/1177729694.

[37] L. Li, J. Yan, H. Wang, and Y. Jin, "Anomaly Detection of Time Series With Smoothness-Inducing Sequential Variational Auto-Encoder," *IEEE Trans. Neural Netw. Learn. Syst.*, vol. 32, no. 3, pp. 1177–1191, Mar. 2021, doi: 10.1109/TNNLS.2020.2980749.

[38] I. J. Goodfellow, Y. Bengio, and A. C. Courville, *Deep Learning*. in Adaptive computation and machine learning. MIT Press, 2016. [Online]. Available: http://www.deeplearningbook.org/

[39] J.-Y. Zhu, T. Park, P. Isola, and A. A. Efros, "Unpaired Image-to-Image Translation Using Cycle-Consistent Adversarial Networks," in *2017 IEEE International Conference on Computer Vision (ICCV)*, Oct. 2017, pp. 2242–2251. doi: 10.1109/ICCV.2017.244.

[40] E. Eldele *et al.*, "Time-Series Representation Learning via Temporal and Contextual Contrasting." arXiv, Jun. 26, 2021. doi: 10.48550/arXiv.2106.14112.

[41] K. He, H. Fan, Y. Wu, S. Xie, and R. Girshick, "Momentum Contrast for Unsupervised Visual Representation Learning," in *2020 IEEE/CVF Conference on Computer Vision and Pattern Recognition (CVPR)*, Jun. 2020, pp. 9726–9735. doi: 10.1109/CVPR42600.2020.00975.

[42] T. Chen, S. Kornblith, M. Norouzi, and G. Hinton, "A Simple Framework for Contrastive Learning of Visual Representations." arXiv, Jun. 30, 2020. doi: 10.48550/arXiv.2002.05709.

[43] D. Djurdjanovic, J. Lee, and J. Ni, "Watchdog Agent—an infotronics-based prognostics approach for product performance degradation assessment and prediction," *Adv. Eng. Inform.*, vol. 17, no. 3, pp. 109–125, Jul. 2003, doi: 10.1016/j.aei.2004.07.005.

[44] "NSL-KDD Datasets | Research." Accessed: Feb. 09, 2023. [Online]. Available: https://www.unb.ca/cic/datasets/nsl.html

[45] M. Tavallaee, E. Bagheri, W. Lu, and A. A. Ghorbani, "A detailed analysis of the KDD CUP 99 data set," *2009 IEEE Symp. Comput. Intell. Secur. Def. Appl.*, pp. 1–6, Jul. 2009, doi: 10.1109/CISDA.2009.5356528.

[46] S. J. Stolfo, W. Fan, W. Lee, A. Prodromidis, and P. K. Chan, "Cost-based modeling for fraud and intrusion detection: results from the JAM project," in *Proceedings DARPA Information Survivability Conference and Exposition. DISCEX'00*, Jan. 2000, pp. 130–144 vol.2. doi: 10.1109/DISCEX.2000.821515.

[47] R. P. Lippmann *et al.*, "Evaluating intrusion detection systems: the 1998 DARPA off-line intrusion detection evaluation," in *Proceedings DARPA Information Survivability Conference and Exposition. DISCEX'00*, Jan. 2000, pp. 12–26 vol.2. doi: 10.1109/DISCEX.2000.821506.

[48] M. Friedman, "The Use of Ranks to Avoid the Assumption of Normality Implicit in the Analysis of Variance," *J. Am. Stat. Assoc.*, vol. 32, no. 200, pp. 675–701, 1937, doi: 10.2307/2279372.

[49] P. B. Nemenyi, "Distribution-free multiple comparisons," 1963.